\documentclass[12pt]{iopart}

\renewcommand{\fnsymbol}[1]{^\arabic{footnote}}

\newcommand*{\mysim}{\mathord{\sim}}
\newcommand{\beq}{\begin{equation}}
\newcommand{\eeq}{\end{equation}}
\newcommand{\beqn}{\begin{eqnarray}}
\newcommand{\eeqn}{\end{eqnarray}}

\newcommand{\sunmass}{M_{\odot}}

\def\eadnew#1#2{\address{#2 E-mail: \mailto{#1}}}

\expandafter\let\csname equation*\endcsname\relax
\expandafter\let\csname endequation*\endcsname\relax
\usepackage{amsmath}
\usepackage{amssymb}

\usepackage{tablefootnote}
\usepackage{enumerate}
\usepackage{ellipsis}
\usepackage{afterpage}
\usepackage{hhline}
\usepackage{multirow}
\usepackage{graphicx}
\usepackage{adjustbox}
\usepackage{epsf}
\usepackage{longtable}
\usepackage{hyperref}
\usepackage{cleveref}[2012/02/15]
\usepackage{graphics,epsfig,placeins,wrapfig}
\usepackage{subcaption}
\usepackage[usenames]{color}
\usepackage{soul}
\usepackage{floatrow}
\usepackage{cite}
\usepackage{braket}
\usepackage{tabularx,booktabs}
\newcolumntype{Y}{>{\centering\arraybackslash}X}
\crefformat{footnote}{#2\footnotemark[#1]#3}

\floatstyle{plaintop}
\restylefloat{table}

\begin{document}

\title{Improving performance of SEOBNRv3 by $\mysim$300x}

\author{
  Tyler D.~Knowles$^{1,*}$,
  Caleb Devine$^{1}$,
  David A.~Buch$^{1}$,\\
  Serdar A.~Bilgili$^{2}$,
  Thomas R.~Adams$^{1}$,
  Zachariah B.~Etienne$^{1,3}$,\\
  Sean T.~McWilliams$^{2,3}$
}

\address{$^{1}$ Department of Mathematics, West Virginia University,
	Morgantown, WV 26506, USA}
\address{$^{2}$ Department of Physics and Astronomy, West Virginia
	University, Morgantown, WV 26506, USA}
\address{$^{3}$ Center for Gravitational Waves and Cosmology, West
	Virginia University, Chestnut Ridge Research Building, Morgantown,
	WV 26505, USA}

\eadnew{tk0014@mix.wvu.edu}{$^{*}$}

\begin{abstract}

When a gravitational wave is detected by Advanced LIGO/Virgo, sophisticated
parameter estimation (PE) pipelines spring into action. These pipelines
leverage approximants to generate large numbers of theoretical gravitational
waveform predictions to characterize the detected signal.
One of the most accurate and physically comprehensive classes of approximants
in wide use is the ``Spinning Effective One Body--Numerical Relativity''
(SEOBNR) family. Waveform generation with these approximants can be
computationally expensive, which has limited their usefulness in multiple data
analysis contexts. In prior work we improved the performance of the
aligned-spin approximant SEOBNR version 2 (v2) by nearly 300x. In this work we
focus on optimizing the full eight-dimensional, precessing approximant SEOBNR
version 3 (v3). While several v2 optimizations were implemented during its
development, v3 is far too slow for use in state-of-the-art source
characterization efforts for long-inspiral detections. Completion of a PE run
after such a detection could take centuries to complete using v3. Here we
develop and implement a host of optimizations for v3, calling the optimized
approximant v3\_Opt. Our optimized approximant is about 340x faster than v3,
and generates waveforms that are numerically indistinguishable.
\end{abstract}

\maketitle

\section{Introduction}

With its first detections of gravitational waves~\cite{Detection1,Detection2,
Detection3,Detection4,Detection5,Detection6}, the Advanced Laser
Interferometer Gravitational Wave Observatory (Advanced LIGO) has provided a
fundamentally new means of observing the Universe. At the heart of
each of these detections was a merger of compact binaries.
In such binaries, each compact object possesses four intrinsic parameters:
mass, and the three components of the spin vector. Inferring all eight
intrinsic parameters\footnote{\textit{Intrinsic} parameters are fundamental to
the underlying physics of the system. In contrast, \textit{extrinsic}
parameters are related to the observer (e.g.~polarization, sky location, and
distance) and are not considered in this paper. Some authors refer to seven
intrinsic parameters in the full-dimensional space, which include each spin
component and the mass ratio of the system. This is because the total mass of
the system is simply a scaling factor; we choose to refer to eight parameters
since the total mass sets the time and frequency scales and therefore must be
considered in PE.} from a gravitational wave observation, which analysis is
part of the more general \textit{parameter estimation} (PE), remains a
challenging and computationally expensive enterprise.

The LIGO/Virgo Scientific Collaboration (LVC) performs PE in a Bayesian
framework, implemented within the \textsc{lalinference} software package that
is part of the larger open-source software framework
\texttt{LALSuite}~\cite{lalsuite}. In such a framework, we sample the
posterior distribution by repeatedly calculating the likelihood that a
particular waveform matches the data and applying Bayes' theorem. Evaluating
the likelihood requires the rapid, sequential generation of as many as 
$\mysim10^8$ theoretical gravitational wave predictions~\cite{Veitch2015}.
Generating so many predictions via a full solution of the general relativistic
field equations (using the tools of numerical relativity) would be far too
computationally expensive. Thus theoretical models adopted for PE generally
employ approximate solutions called \emph{approximants}. State-of-the-art
approximants adopt post-Newtonian techniques for evaluating the gravitational
waveform throughout most of the inspiral and ringdown, and inject information
from numerical relativity calculations for the late inspiral and merger.

One such gravitational wave approximant is the Spinning Effective One
Body--Numerical Relativity (SEOBNR) algorithm. This algorithm marries an
effective-one body inspiral gravitational waveform approximation---with
unknown higher-order terms fit to numerical relativity-generated gravitational
wave predictions---to a black hole ringdown model~\cite{Pan2014}. In
particular, SEOBNR starts with the Effective One Body (EOB) approach to
non-spinning binary modeling~\cite{Buonanno1999} by mapping the dynamics of
the two-body system to the dynamics of an effective particle moving in a
deformed Schwarzschild metric. This work was then extended to include the
effects of spinning, precessing binaries~\cite{Buonanno2006}. Implemented
numerically, this Spinning EOB procedure adopts a precessing source frame in
which precession-induced variations in amplitude and phase are minimized
during inspiral, and a source frame aligned with the spin of the final body
for matching the inspiral to the merger-ringdown~\cite{Pan2014}. 

The other widely adopted approximant within the LVC for PE is the Phenom
series of phenomenological waveform models. These waveform models are based on
the combination of accurate post-Newtonian inspiral models with late-inspiral
and merger phenomenological fits to suites of numerical relativity
simulations~\cite{Santamaria2010}. More recently, Phenom models have been 
built to include the effects of precession~\cite{Hannam2014}. In particular,
precession effects are included by using post-Newtonian methods to compute
precession angles and then ``twisting'' the underlying non-precessing
model~\cite{Hannam2014,Husa2016,Khan2016}. Phenom models are simulated
completely in the frequency domain, and therefore simplify some aspects of
analysis. The only Phenom model designed to generate gravitational waveform
predictions across all eight dimensions of parameter space is
PhenomP~\cite{Hannam2014}, which was extensively used in the first six
detection papers. We remark that Phenom is limited to a relatively small
number of numerical relativity simulations against which it has been
calibrated, and it is difficult to determine the degree of systematic
uncertainty in the model without appealing to another model for comparison. 

Evaluating the systematic uncertainties of the Phenom model requires
construction of an independent gravitational waveform model with independent
systematics, and the SEOBNR family of models is a good candidate for this
task. The only SEOBNR model capable of generating theoretical gravitational
waveform predictions in all 8 intrinsic dimensions of parameter space is the
third version of the model, v3; the first and second versions were restricted
to aligned-spin cases. In particular, v3 was built to accommodate arbitrary
mass ratios, spin magnitudes, and spin orientations and has been calibrated and
validated against a variety of numerical relativity
simulations~\cite{Babak2017}. Thus v3 is vital for precessing compact binary
merger PE.

Unfortunately, v3 is too currently too slow for PE. A single waveform
generation across the LIGO band for, say, a black hole-neutron star system
using v3 can take as long as an hour on a modern desktop computer. If LIGO
observed a black hole-neutron star system merge, a
sequential-gravitational-wave-generation PE would take thousands of years.
Attempts to overcome the computational challenge of generating such
time-consuming gravitational waveforms include the construction of Reduced
Order Model (ROM) approximants. ROMs make use of multidimensional
interpolations between sampled points in another underlying approximant. For
example, a ROM based on the aligned-spin SEOBNR version 2 (v2)
approximant~\cite{Purrer2016,Field2014} is constructed by first generating an
extensive collection of waveform predictions using v2 that adequately samples
the 4D parameter space reliably covered by v2. Then to obtain the
gravitational waveform at any desired point in parameter space, the ROM simply
interpolates within the four dimensions of sampled parameter space. A ROM
version of v2 can generate waveforms up to $\mysim3000$x faster than v2
directly~\cite{Purrer2016}, which explains in part why ROMs enjoy such
widespread use within the LVC for data analysis applications. 

While ROMs have been constructed with favorable performance characteristics in
aligned-spin situations, the cost of generating a ROM grows exponentially with
the dimension of the ROM (though see~\cite{Field2012} for ideas on combating
this using a reduced basis approach). No strategy yet exists that can perform
the 8-dimensional (8D) interpolations faster than the 8D approximant; until
such a strategy is invented, the most promising way to improve the performance
of theoretical waveform generation in the full 8D parameter space will be to
optimize the approximant directly. As a proof-of-principle, we demonstrated
that such an approach is capable of improving the performance of the
aligned-spin v2 approximant by a typical factor of $\mysim280$x~\cite{v2opt}.
We call our optimized v2 approximant v2\_opt. The precessing (8D) v3
approximant was in development as we independently prepared v2\_opt, and thus
originally contained all the same inefficiencies as v2. This suggests that if
the full suite of optimizations we implemented in v2 were incorporated into
v3, v3-based PE timescales might drop by two orders of magnitude at least.

This paper documents our incorporation of applicable v2 optimizations into v3,
as well as our implementation of innovative new optimization ideas, which
together act to speed up v3 by $\mysim340$x. Optimization strategies are
summarized in Sec.~\ref{OptStrats}. Section~\ref{Results}  presents code
validation tests that demonstrate roundoff-level agreement between v3 and our
latest optimized version of v3, designated v3\_Opt, along with benchmarks
providing an overview of performance gains across parameter space in v3\_Opt.
{\bf For convenience, Table~\ref{approx_conv} defines all SEOBNR approximants
referenced in this paper.}

  \begin{table}
    \begin{adjustbox}{max width=\textwidth}
    \centering
      \begin{tabular}{|c|c|l|}
        \hline
        \textbf{Base} & \textbf{Approx.} & \textbf{Description} \\
        \textbf{Approximant}  & \textbf{Name} & \\
        \hline\hline
        SEOBNRv2 & v2 & Initial SEOBNRv2 implementation\tablefootnote{\label{2cce415}As of publication, the most recent updates to v2/v2\_opt are found on commit ID \texttt{2cce415} in the \texttt{LALSuite} \texttt{master} branch.}; see \cite{Taracchini2014}. \\\cline{2-3}
        (spin-aligned) & v2\_opt & Optimized v2\cref{2cce415}; see \cite{v2opt}. \\
        \hline
        SEOBNRv3 & v3\_preopt & Initial SEOBNRv3 implementation\tablefootnote{To generate a waveform with v3\_preopt, download \texttt{LALSuite} from the archived repository page \url{https://git.ligo.org/lscsoft/lalsuite-archive/tree/14414694698a2f18c9135445003cade805ad2096} and use approximant tag SEOBNRv3.}; see \cite{Pan2014}. \\\cline{2-3}
        (precessing) & v3 & Partially optimized v3\_preopt with bug fixes\tablefootnote{\label{19e95b4}As of publication, the most recent updates to v3 and v3\_opt are found on commit ID \texttt{19e95b4} in the \texttt{LALSuite} \texttt{master} branch.}. \\\cline{2-3}
        & v3\_pert & v3 with machine-$\epsilon$ mass perturbation \cref{19e95b4}. \\\cline{2-3}
        & v3\_opt &v3 optimized similarly to v2\_opt \cref{19e95b4}. \\\cline{2-3}
        & v3\_Opt & v3\_opt with new optimization strategies\tablefootnote{\label{Opt}Approximants v3\_opt and v3\_opt\_rk4 were updated to run v3\_Opt and v3\_Opt\_rk4, respectively, on commit ID \texttt{1391f77} in the \texttt{LALSuite master} branch.}. \\\cline{2-3}        
        & v3\_Opt\_rk4 & v3\_Opt implementing RK4 rather than RK8\cref{Opt}. \\
        \hline
      \end{tabular}
      \caption{Approximant naming conventions. These conventions apply
        throughout this paper.}\label{approx_conv}
    \end{adjustbox}
  \end{table}

\section{SEOBNRv3\_opt: Optimizations migrated from v2\_opt}
\label{OptStrats}
Optimizations to v3 were performed in two phases. In the first phase,
described in Sec.~\ref{v2opts}, we migrated to v3 all applicable optimizations
developed during the preparation of v2\_opt. Sections~\ref{v3opts:gad} and~\ref{v3opts:do} detail the second phase of optimization, outlining new strategies
incorporated into v3\_Opt.

\subsection{Migrated Optimizations}
\label{v2opts}
Here we summarize the optimizations to v2 which were migrated to v3 and thus
implemented in v3\_opt.

\begin{itemize}
  \item \emph{Switching compilers}. Switching from the \texttt{GNU Compiler
    Collection} (\texttt{gcc})~\cite{gcc} \texttt{C} compiler to the
    \texttt{Intel Compiler Suite} (\texttt{icc})~\cite{icc} \texttt{C}
    compiler improves performance by roughly a factor of 2x. It is well-known
    that the \texttt{icc} compiler often produces more efficient executables
    than the \texttt{gcc} compiler\footnote{We used the following compiler
    flags when compiling with \texttt{icc}: \texttt{-xHost}, \texttt{-O2}, and
    \texttt{-fno-strict-aliasing}.}.

  \item \emph{Minimize transcendental function evaluations}. The EOB
    Hamiltonian equations of motion were hand-optimized by minimizing calls to
    some expensive transcendental functions such as \texttt{exp()},
    \texttt{log()}, and \texttt{pow()}.

  \item \emph{Replacing finite difference with exact derivatives}. When
    solving the EOB Hamiltonian equations of motion, v3 computes partial
    derivatives of the Hamiltonian using finite difference approximations. We
    replaced these with exact, Mathematica-generated expressions for the
    derivatives, using Mathematica's code generation facilities---which
    includes common subexpression elimination (CSE)---to generate the
    \texttt{C} code~\cite{Mathematica}. Although this alone acts to
    significantly speed up v3, in this work we further optimize these
    Mathematica-generated derivatives.

  \item \emph{Increasing the order of the ODE solver}. v3 solves the EOB
    Hamiltonian equations of motion via a Runge-Kutta fourth order (RK4) ODE
    solver. After implementing exact derivatives, we noticed that the number
    of RK4 steps needed dropped significantly---presumably due to the
    effective removal of high numerical noise intrinsic to finite-difference
    derivatives. We then found that adopting a Runge-Kutta eighth order (RK8)
    ODE solver resulted in 2x larger timesteps, so an even larger speed-up was
    observed.

  \item \emph{Reducing orbital angular velocity calculations}. The orbital
    angular velocity $\omega$ was calculated for each ($\ell,m$) mode (as
    defined in \cite{Pan2014}) inside the ODE solver. As $\omega$ exhibits no
    dependence on $\ell$ or $m$, this expensive recalculation was unnecessary
    and needs only be performed once.
\end{itemize}

For more details about these optimizations see our v2 optimization
paper~\cite{v2opt}. 

\subsection{Guided Automatic Differentiation: A more efficient way of
  generating symbolic derivatives of the Hamiltonian}\label{v3opts:gad}

After migrating the v2 optimizations described in Sec.~\ref{v2opts} to v3,
profiling analyses indicated that approximately $75\%$ of v3\_opt's total
runtime was spent computing the v3 Hamiltonian~\cite{Pan2014} and its partial
derivatives with respect to the twelve degrees of freedom (consisting of three
spatial degrees $\{x,y,z\}$, three momentum degrees $\{p_x,p_y,p_z\}$, and
three spin degrees for each of the two binary components $i\in\{1,2\}$:
$\{s_{i}^{x},s_{i}^{y},s_{i}^{z}\}$).

In v3, the ODE solver computes these partial derivatives by direct evaluations
of the Hamiltonian itself via finite difference
techniques~\cite{Taracchini2014}. In v3\_opt, these numerical derivatives were
replaced with Mathematica-generated exact derivatives. Although these exact
derivatives unlock significant performance gains, the Mathematica-generated
\texttt{C} code was neither particularly human-readable (comprising thousands
of lines of code output by Mathematica's CSE routines) nor particularly
well-optimized (common patterns were still visible and recomputed in the
\texttt{C} code). Attempts to gain performance through consolidation of all
derivatives---as was possible in our optimizations of v2---proved beyond
Mathematica's capabilities when differentiating the v3 Hamiltonian on our
high-performance workstations. Therefore, \texttt{C} codes for all twelve
exact derivatives needed to be output separately, resulting in a significant
number of unnecessary re-computations.

We present here our new strategy for computing partial derivatives of the
Hamiltonian, called \textit{guided automatic differentiation} (GAD), which
results in a significant reduction in computational cost while ensuring the
resulting code is highly human-readable.  GAD is based on forward accumulation
automatic differentiation, with the advantage of the subexpressions being
chosen by hand to minimize the overall number of floating point operations.

The following describes the process of computing a partial derivative of the
v3 Hamiltonian $H$ with respect to an \textit{arbitrary} independent variable
$x_{1}$ using GAD. We may write $H$ in the following form, where $I$ is a set
of input quantities:
\begin{align*}
v_{1} &= f_{1}(I) \\
v_{2} &= f_{2}(v_1,I) \\
v_{3} &= f_{3}(v_1,v_2,I) \\
 &\ \ \vdots  \\
H &= f_{N}(v_{1},v_{2},v_{3},...,I).
\end{align*}
Here $f_\ell$ is the $\ell$th function of the set of input quantities $I$ and
previously computed subexpressions
$\left\{v_{0},v_{1},\hdots,v_{\ell-1}\right\}$. Although $N\approx 200$ for
v3, \textit{for the sake of example} we suppose $N=3$, $I=\{x_{1},x_{2}\}$,
and
\begin{align*}
v_1 &= \sqrt{x_{1}} + ax_{1} \\
v_2 &= \sqrt{x_{2}} + ax_{2} \\
v_3 &= (v_1 + v_2)/(v_1v_2)  \\
H  &= v_3^2.
\end{align*}
We demonstrate GAD by taking a partial derivative of $H$ with respect to the
independent input variable $x_{1}$. Table \ref{doGAD} displays the evolution
of this example code under the GAD scheme, which proceeds as follows:
\begin{enumerate}[1.]
  \item We begin with a list of variables and subexpression computations for
    the Hamiltonian, and translate this \texttt{C} code into the Mathematica
    language.
  \item We parameterize the terms of each subexpression according to their
    dependence on $x_{1}$.
  \item Mathematica computes derivatives of each subexpression.
  \item We convert the Mathematica output into \texttt{C} code.
  \item We replace each occurrence of $x_{1}'$ with 1 and remove terms equal to 0.
\end{enumerate}

The resulting \texttt{C} code is short, optimized, and human readable.
Furthermore, any terms that are common to all derivative expressions are
computed and saved before computing the partial derivatives, further reducing
the computational cost.

\begin{table}
  \centering
  \begin{adjustbox}{max width=\textwidth}
    \begin{tabular}{|l|l|}
      \hline
      \textbf{Step 1: Convert \texttt{C} to Mathematica.} & \textbf{Step 2: Parameterize subexpressions.} \\
        \texttt{v1 = Sqrt[x1] + a*x1}                     & \texttt{v1 = Sqrt[x1[x]] + a*x1[x]}           \\
        \texttt{v2 = Sqrt[x2] + a*x2}                     & \texttt{v2 = Sqrt[x2] + a*x2}                 \\
        \texttt{v3 = (v1 + v2)/(v1*v2)}                   & \texttt{v3 = (v1[x] + v2[x])/(v1[x]*v2[x])}   \\
        \texttt{H~ = v3*v3}                               & \texttt{H = v3[x]*v3[x]}                      \\
      \hline
        \multicolumn{2}{|l|}{\textbf{Step 3: Utilize Mathematica to compute derivatives.}}                 \\
        \multicolumn{2}{|l|}{\texttt{v1' = x1'[x]/(2*Sqrt[x1[x]]) + a*x1'[x]}}                             \\
        \multicolumn{2}{|l|}{\texttt{v2' = 0}}                                                             \\
        \multicolumn{2}{|l|}{\texttt{v3' = (v1[x]*v2[x]*(v1'[x] + v2'[x])-((v1[x] + v2[x])*(v1'[x]*v2[x]}} \\
        \multicolumn{2}{|l|}{\texttt{~~~~~ + v1[x]*v2'[x]))/(v1[x]*v1[x]*v2[x]*v2[x])}}                    \\
        \multicolumn{2}{|l|}{\texttt{H'~ = 2*v3'[x]*v3[x]}}                                                \\
      \hline
        \multicolumn{2}{|l|}{\textbf{Step 4: Convert Mathematica to \texttt{C}; prime notation becomes a protected \texttt{prm} suffix.}} \\
        \multicolumn{2}{|l|}{\texttt{v1prm = x1prm/(2*sqrt(xi)) + a*x1prm}}                                                               \\
        \multicolumn{2}{|l|}{\texttt{v2prm = 0}}                                                                                          \\
        \multicolumn{2}{|l|}{\texttt{v3prm = (v1*v2*(v1prm + v2prm)-((v1 + v2)*(v1prm*v2 + v1*v2prm))/(v1*v1*v2*v2)}}                     \\
        \multicolumn{2}{|l|}{\texttt{Hprm~ = 2*v3prm*v3}}                                                                                 \\
      \hline
        \multicolumn{2}{|l|}{\textbf{Step 5: Replace \texttt{x1prm} with 1 and remove terms equaling 0.}} \\
        \multicolumn{2}{|l|}{\texttt{v1prm = 1./(2*sqrt(x1)) + a}}                                        \\
        \multicolumn{2}{|l|}{\texttt{v3prm = (v1*v2*v1prm-(v1 + v2)*v1prm*v2)/(v1*v1*v2*v2)}}             \\
        \multicolumn{2}{|l|}{\texttt{Hprm~ = 2*v3prm*v3}}                                                 \\
      \hline
    \end{tabular}
  \end{adjustbox}
  \caption{Step-by-step GAD code evolution.}\label{doGAD}
\end{table}

Since each $v_{\ell}$ is merely an intermediate of $H$, there is significant
freedom in our choice of the set of subexpressions
$\mathcal{V}\equiv\left\{v_{1},v_{2},\hdots,v_{N-1}\right\}$. Our choices do,
however, have a direct effect on the number of calculations necessary to
compute $\partial_{x_{1}}H$, which we measure in floating point operations
(FLOPs\footnote{Not to be confused with ``FLOPs per second'' (FLOPS).}.) Our
goal in GAD, therefore, is to choose $\mathcal{V}$ to minimize the number of
FLOPs needed to compute $\partial_{x_{1}}H$.

In general, the largest contributor to FLOPs is the product rule. If there are
$M$ different subexpressions multiplied together in a given expression,
computing the derivative will require $\mathcal{O}(M^2)$ FLOPs. If we
therefore choose $\mathcal{V}$ such that each $v_{\ell}$ contains no more than
two previous subexpressions multiplied together, we should minimize the
overall cost. We expect a significant reduction in FLOPs to correspond to a
significant reduction in the time to generate a waveform.

We estimated the number of FLOPs based on benchmarks provided
in~\cite{FLOPsite} for CPUs corresponding to the CPU family in our
workstations (Intel Core i7-6700) and generated Table \ref{countratio}. We
emphasize that the values listed in Table~\ref{countratio} are truly rough
estimates, used only to provide us general direction as we seek an optimal
$\mathcal{V}$.

\begin{table}
  \centering
  \begin{adjustbox}{max width=\textwidth}
    \begin{tabular}{|c|c|c|c|c|c|c|c|}
      \hline
      $a+b$ & $a-b$ & $a*b$ & $a=b$ & $a/b$ & ${\tt sqrt}(a)$ & ${\tt log}(a)$ & ${\tt pow}(a,b)$  \\ \cline{3-5}
      \hline
        1   &   1   &   1   &   1   &   3   &     3     &    24    &     24      \\ \cline{3-5}
      \hline
    \end{tabular}
  \end{adjustbox}
  \caption{Relative FLOPs count of the mathematical
    operations.}\label{countratio}
\end{table}

Table~\ref{FLOPs} compares the number of Hamiltonian derivative FLOPs under
GAD to the number in the exact derivatives (EDs) generated by Mathematica's
CSE code generation algorithm.  In principle, the difference in FLOPs between
ED and GAD schemes may be used to predict the waveform generation speedup
factor. A direct comparison from Table~\ref{FLOPs} indicates a 3.6x reduction
in FLOPs when using GAD. For a double neutron star coalescence, Hamiltonian
derivative computations constitute about 80\% of waveform generation time.
This suggests a speedup factor of 2.3x. Waveform generation times for three
scenarios comparing ED and GAD implemented in v3\_opt are shown in
Table~\ref{previewbench}, and demonstrates a speedup factor of about 1.7x. We
emphasize again that counting FLOPs using the relative values of
Table~\ref{countratio} only provides a rough estimate of the reduction in
FLOPs, and the compiler itself rearranges arithmetic expressions to minimize
FLOPs as well so the gap between our estimated and observed speed-ups is not
surprising.

  \begin{table}
    \centering
    \begin{adjustbox}{max width=\textwidth}
      \begin{tabular}{|c|ccc|c|}
        \hline
        Derivative scheme & Space derivative &  Momentum derivative &  Spin derivative &  Total   \\
                          & (FLOPs)          & (FLOPs)              & (FLOPs)          &  (FLOPs) \\
        \hline
        ED                & 3 x 5073         & 3 x 2319             &  6 x 4333        &  48174   \\
        GAD               & 3 x 1418         & 3 x 527              &  6 x 1264        &  13419   \\
        \hline
      \end{tabular}
    \end{adjustbox}
    \caption{Number of FLOPs using ED versus GAD methods.}\label{FLOPs}
  \end{table}

  \begin{table}
    \centering
    \begin{adjustbox}{max width=\textwidth}
      \begin{tabular}{|c|c|c|c|}
        \hline
                                                                            & v3\_opt (s)       & v3\_opt (s)       \\
        \textbf{Parameters}                                                 & ED                & GAD               \\
        \hline
        \textbf{Neutron Star Binary}                                        & 36.75             & 20.49             \\
        $1.4\sunmass + 1.4\sunmass$, $s_{1}^{y} = 0.05$                     &                   & x(\textbf{1.79})  \\ 
        \hline
        \textbf{Black Hole + Neutron Star}                                  & 8.07              & 4.69              \\
        $10\sunmass + 1.4 \sunmass$, $s_{1}^{y} = 0.4$                      &                   & x(\textbf{1.72})  \\
        \hline
        \textbf{Black Hole Binary} (GW150914-like)                          &                   &                   \\
        $36\sunmass + 29\sunmass$                                           & 0.64              & 0.38              \\
        $s_{1}^{y}=0.05,\ s_{1}^{z}=0.5,\ s_{2}^{y}=-0.01,\ s_{2}^{z}=-0.2$ &                   & x(\textbf{1.68})  \\
        \hline
      \end{tabular}
    \end{adjustbox}
    \caption{Benchmark comparison of ED to GAD strategies. In each
      scenario, we adopt a 10Hz start frequency.}\label{previewbench}
  \end{table}

\subsection{Dense Output: A more efficient way of interpolating
  sparsely-sampled data}
\label{v3opts:do}

An RK4 ODE solver with adaptive timestep control solves the EOB Hamiltonian
equations of motion in v3; thus solutions are unevenly sampled in time.
Subsequent analyses require mapping these data into the frequency domain via
the fast Fourier transform (FFT), which expects evenly-sampled data. Rather
than restricting the integration timestep, v3 uses cubic splines to
interpolate the Hamiltonian solutions after RK4 runs to completion. During
optimization of v2, the GSL cubic spline interpolation routine was optimized
and gave significant performance gains. During optimization of v3, it was
discovered that third-order Hermite interpolation made v3\_Opt more faithful
to v3 (see Section \ref{Results}). Hermite interpolation requires only two
function values and the derivatives at those values, which are available at
each step of RK8. Thus we may interpolate the sparsely-sampled data to the
desired evenly-sampled data ``on the fly'' during integration. Such an
integration routine is called a \textit{dense output} method~\cite{NR}. In
particular, suppose the RK8 integrator computes the solution $y(t_{0})$ and
$y(t_{1})$ at times $t_{0}$ and $t_{1}$ with timestep $h$ and derivative
values $y'_{0}$ $y'_{1}$. Then for any $0\le\theta\le1$, we have
  \begin{equation*}
  \begin{adjustbox}{max width=\textwidth}
      $y(t_{0} + \theta h) = (1-\theta)y(t_{0}) + \theta y(t_{1}) + \theta(\theta-1)\left[(1-2\theta)(y(t_{1}) - y(t_{0}) + (\theta-1)hy'(t_{0}) + \theta hy'(t_{1})\right]$.
  \end{adjustbox}
  \end{equation*}
As this cubic Hermite interpolation routine uses both the solution data and
derivative values at each point, it therefore requires only the output of the
RK8 integration and no further data storage or function evaluations.

\section{Results}\label{Results}
In Sec.~\ref{subsec:faith} we establish that v3\_Opt produces waveforms which
agree with v3 at the level of roundoff error. Section \ref{subsec:bench} then
describes the process of measuring speedup and demonstrates the speedup factor
achieved.

\subsection{Determining Faithfulness}\label{subsec:faith}
Given two waveforms $h_1(t)$ and $h_2(t)$ (in the time domain), we determine
if $h_{1}(t)$ is \emph{faithful} to $h_{2}(t)$ using the LVC's open-source
\textsc{PyCBC} software~\cite{pycbc-software,Canton2014,Usman2015}. This
computation depends on the following definitions, which we write in the same
form as~\cite{v4}. The \emph{noise-weighted overlap} between $h_{1}$ and
$h_{2}$ is defined as
  \[
    (h_1 | h_2) \equiv 4{\rm Re} \int_{f_{l}}^{f_{h}} \frac{ \tilde{h}_{1}(f)\tilde{h}^{*}_{2}(f) }{ S_{n}(f) } {\rm d}f
  \]
with $\tilde{h}_{i}(f)$ denoting the Fourier transform of the waveform
$h_{i}(t)$, $h_{i}^{*}$ denoting the complex conjugate of $h_{i}$, $f_{l}$ and
$f_{h}$ denoting the endpoints of the range of frequencies of interest, and
$S_n(f)$ denoting the one-sided power spectral density (PSD) of the LIGO
detector noise. We chose $f_{l}=20$ Hz and $S_{n}(f)$ to be Advanced LIGO's 
design zero-detuned high-power noise PSD~\cite{noisecurve}. For each waveform,
$f_{h}$ is the Nyquist critical frequency~\cite{NR}. We then define the
\emph{faithfulness} between $h_{1}$ and $h_{2}$ to  be the overlap between the
normalized waveforms maximized over relative time and phase shifts:
  \[
    \braket{h_{1} | h_{2}} \equiv \max_{\phi_c, t_c} \frac{ (h_1(\phi_c,t_c) | h_2) }{ \sqrt{(h_1 | h_1)(h_2 | h_2)} }.
  \]
Here $t_{c}$ and $\phi_c$ denote the coalescence time and phase, respectively.
Note that normalization forces $\braket{h_{1} | h_{2}} \in [0,1]$, with
$\braket{h_{1} | h_{2}} = 1$ indicating complete overlap (and therefore a
perfectly faithful waveform) while $\braket{h_{1} | h_{2}} = 0$ indicates no
overlap (an \emph{unfaithful} waveform\footnote{Another common measure in
faithfulness tests is \textit{mismatch}, defined as
$1-\braket{h_{1} | h_{2}}$.}). For each faithfulness test conducted, we
generate a waveform with two different approximants and the same set of input
parameters.

We ran 100,000 faithfulness tests for each set of waveform approximants we
wished to compare. The input parameters for each test are randomly chosen by
\textsc{PyCBC} with bounds as outlined in Table \ref{faithparams}; these
bounds are chosen to capture the relevant parameter space for v3.  Note that
each of the spin parameters $s_{i}^{x}, s_{i}^{y}, s_{i}^{z}$ are chosen
randomly in $(-1,1)$ with the constraint
\[
  \sqrt{ (s_{i}^{x})^{2} + (s_{i}^{y})^{2} + (s_{i}^{z})^{2} } \le 0.99, \ \ i \in \{ 1,2\}.
\]

\begin{table}
  \begin{tabular}{|c|c|}
    \hline
    Mass of Object 1 (solar masses)            & $m_{1}\in[1,100]$               \\
    \hline
    Mass of Object 2 (solar masses)            & $m_{2}\in[1,100]$               \\
    \hline
    Spin magnitude of Object 1 (dimensionless) & $\lvert a_{1}\rvert\in[0,0.99]$ \\
    \hline
    Spin magnitude of Object 2 (dimensionless) & $\lvert a_{2}\rvert\in[0,0.99]$ \\
    \hline
    Binary total mass (solar masses)           & $m_{\rm total}\in[4,100]$       \\
    \hline
    Starting orbital frequency (Hz)            & $f=19$                          \\
    \hline
  \end{tabular}
\caption{Ranges of values for random input parameters in our faithfulness
  tests.}\label{faithparams}
\end{table}

The specific faithfulness runs we conducted were organized as follows. The
approximant v3\_pert is identical to v3 except $m_{1}$ is replaced with
$m_{1}\left( 1 + 10^{-16} \right)$; such a perturbation should result
in waveforms that are nearly identical and provides a measure of how sensitive
v3 is to roundoff error. Thus faithfulness tests comparing v3 and v3\_pert
provide a ``control'' against which we compare the faithfulness of v3\_Opt to
v3. As another point of comparison, we also test v3 (which is RK4-based)
against the RK4-based v3\_Opt\_rk4. For each approximant comparison we
compare the effect of increasingly stricter ODE solver tolerance. By default,
v3 sets the ODE solver's absolute and relative error tolerances to
$\varepsilon\equiv1\times10^{-8}$; we compare faithfulness at tolerances of
 $\varepsilon$, $\varepsilon\times10^{-1}$, $\varepsilon\times10^{-2}$,
$\varepsilon\times10^{-3}$, and $2\varepsilon\times10^{-4}$. Finally, we also
consider the effect of compiler choice on faithfulness and so conduct
faithfulness runs using both \texttt{gcc} and \texttt{icc}. Table
\ref{faithresults} summarizes the faithfulness tests conducted and their
results; the rightmost column displays the counting error $\sqrt{n}$ for the
number of waveforms $n$ with $\braket{\cdot|\cdot} < 0.999$.

\begin{table}
\begin{adjustbox}{max width=\textwidth}
\begin{tabularx}{1.3\textwidth}{c c c *{5}{Y} c}
\toprule
                                  &               & \bf{ODE}                   & \multicolumn{5}{c}{\textbf{Number of waveforms with faithfulness}} & \textbf{Counting} \\
\cmidrule(lr){4-8}
\bf{Comparison}                   & \bf{Compiler} & \bf{tolerance}             & $<0.8$ & $<0.9$ & $<0.95$ & $<0.99$ & $<0.999$                     & \textbf{Error} \\
\midrule
v3 vs.~v3\_pert                   & \texttt{gcc}  & $\varepsilon$              & 1      & 5      & 13      & 104     & 399                          & $\pm20$ \\
(per $10^{5}$ for $10^{6}$ tests) & \texttt{icc}  & $\varepsilon$              & 1.0    & 4.2    & 11.5    & 109.0   & 398.2                        & $\pm6.3$ \\
\hline
v3 vs.~v3\_Opt                    & \texttt{gcc}  & $\varepsilon$              & 5      & 28     & 136     & 1184    & 5466                         & $\pm74$ \\
                                  & \texttt{icc}  & $\varepsilon$              & 5      & 28     & 135     & 1174    & 5509                         & $\pm74$ \\
                                  & \texttt{icc}  & $\varepsilon\times10^{-1}$ & 2      & 16     & 44      & 327     & 1510                         & $\pm39$ \\
                                  & \texttt{icc}  & $\varepsilon\times10^{-2}$ & 0      & 2      & 12      & 143     & 727                          & $\pm27$ \\
                                  & \texttt{icc}  & $\varepsilon\times10^{-3}$ & 1      & 3      & 8       & 80      & 511                          & $\pm23$ \\
                                  & \texttt{icc}  & $2\varepsilon\times10^{-4}$& 1      & 1      & 2       & 60      & 457                          & $\pm21$ \\
\hline
v3 vs.~v3\_Opt\_rk4               & \texttt{gcc}  & $\varepsilon$              & 1      & 9      & 35      & 427     & 1529                         & $\pm39$ \\
                                  & \texttt{icc}  & $\varepsilon$              & 0      & 9      & 35      & 420     & 1510                         & $\pm39$ \\
                                  & \texttt{icc}  & $\varepsilon\times10^{-1}$ & 1      & 7      & 24      & 223     & 926                          & $\pm30$ \\
                                  & \texttt{icc}  & $\varepsilon\times10^{-2}$ & 0      & 0      & 8       & 114     & 585                          & $\pm24$ \\
                                  & \texttt{icc}  & $\varepsilon\times10^{-3}$ & 1      & 3      & 8       & 77      & 483                          & $\pm22$ \\
                                  & \texttt{icc}  & $2\varepsilon\times10^{-4}$& 1      & 2      & 3       & 52      & 423                          & $\pm21$ \\
\bottomrule
\end{tabularx}
\end{adjustbox}
\caption{Summary of \texttt{PyCBC} faithfulness results. Here
$\varepsilon=1\times10^{-8}$ and each row reports the results of a run of
\textbf{100,000 faithfulness tests}. \texttt{icc} refers to Intel compiler
version 15.5.223, while \texttt{gcc} refers to GNU compiler version
4.9.}\label{faithresults}
\end{table}

We comment on the values in Table \ref{faithresults}. For a couple of
parameters for which $\braket{\cdot | \cdot} < 0.8$ when comparing v3 to
v3\_Opt compiled with \texttt{gcc}, one author back-traced a significant
difference between v3 and v3\_Opt to the ODE stopping condition or the time of
maximum amplitude being clearly wrong in v3 but not v3\_Opt. In particular,
there are some algorithms within v3 that are fundamentally non-robust, and
v3\_Opt inherits most of these functions. The RK8 integration of v3\_Opt
should be just as accurate as the RK4 integration of v3\_Opt\_rk4 when the
tolerances are equal, but the output from RK8 should be much sparser (by more
than a factor of 2) than RK4. Since we observe worse faithfulness with v3\_Opt
than v3\_Opt\_rk4, we conclude that most of the truncation error stems from
the interpolation of the sparsely-sampled ODE solution to a uniform timestep.

Most importantly, notice that as we make the ODE solver's tolerance $\varepsilon$
stricter (resulting in smaller errors and more finely sampled output data from
the ODE solver), the faithfulness between v3 and v3\_Opt improves to the level
of agreement between v3 and v3\_pert.  Thus we conclude that v3\_opt generates
roundoff-level agreement in the limit of $\varepsilon\to0$ with errors dominated
by interpolation otherwise.

\subsection{Performance Benchmarks}\label{subsec:bench}
In order to capture the full effect of our optimizations to v3, we compared
waveform generation times of v3\_Opt with waveform generation times of
v3\_preopt. In particular, v3\_preopt lacks by-hand optimizations of the EOB
Hamiltonian implemented in the development of v2\_opt; thus unnecessary
computations of transcendental functions \texttt{pow()}, \texttt{log()}, and
\texttt{exp()} remain therein. All reported benchmarks were completed on a
single core of a modern desktop computer with an Intel Core i7-7700 CPU and 64
GB RAM.

To highlight cases of interest, Table \ref{benchmarkresults} summarizes
benchmarks of v3\_Opt and v3\_Opt\_rk4 in comparison to v3\_preopt for a
handful of scenarios of interest to LIGO. The speedup factors are also
included, with speedup simply defined to be the ratio of time to generate a
waveform with v3\_preopt to the time to generate the same waveform with
v3\_Opt or v3\_Opt\_rk4.

  \begin{table}
    \centering
    \begin{adjustbox}{max width=\textwidth}
      \begin{tabular}{|c|c|c|c|c|}
        \hline
        Physical scenario                   & v3\_preopt                    & v3\_Opt\_rk4     & v3\_Opt                   & v3\_Opt                  \\
                                            & \texttt{gcc}, (s)                  & \texttt{gcc}, (s)               & \texttt{gcc}, (s)               & \texttt{icc}, (s)                \\
        \hline
        \hline
        DNS, $s_{2}^{y}=0.05$               & \multirow{2}{*}{8618.60}           & 98.51                           & 42.85                           & 21.22                            \\ \cline{3-5}
        $1.3\sunmass+1.3\sunmass$           &                                    & x(\textbf{87.49})               & x(\textbf{201.1})               & x(\textbf{406.2})                \\
        \hline
        BHNS, $s_{\rm NS}^{y}=0.05$         & \multirow{2}{*}{2760.77}           & 20.75                           & 8.84                            & 4.37                             \\ \cline{3-5}
        $10\sunmass+1.3\sunmass$            &                                    & x(\textbf{133.0})               & x(\textbf{312})                 & x(\textbf{632})                  \\
        \hline
        BHB, $s_{2}^{y}=0.05$               & \multirow{2}{*}{127.71}            & 1.70                            & 0.90                            & 0.46                             \\ \cline{3-5}
        $16\sunmass+16\sunmass$             &                                    & x(\textbf{75.1})                & x(\textbf{140})                 & x(\textbf{280})                  \\
        \hline
        BHB, $s_{1}^{y}=s_{2}^{y}=0.9$      & \multirow{2}{*}{168.13}            & 1.75                            & 0.91                            & 0.46                             \\ \cline{3-5}
        $16\sunmass+16\sunmass$             &                                    & x(\textbf{96.1})                & x(\textbf{180})                 & x(\textbf{370})                  \\
        \hline
        BHB, $s_{1}^{y}=s_{2}^{z}=0.9$      & \multirow{2}{*}{235.53}            & 3.48                            & 1.55                            & 0.76                             \\ \cline{3-5}
        $10\sunmass+10\sunmass$             &                                    & x(\textbf{67.7})                & x(\textbf{152})                 & x(\textbf{310})                  \\
        \hline
        BHB, GW150914-like                  & \multirow{4}{*}{31.48}             & \multirow{2}{*}{0.75}           & \multirow{2}{*}{0.51}           & \multirow{2}{*}{0.27}            \\
        $36\sunmass+29\sunmass$             &                                    &                                 &                                 &                                  \\ \cline{3-5}
        $s_{1}^{y}=0.05$, $s_{1}^{z} = 0.5$ &                                    & \multirow{2}{*}{x(\textbf{42})} & \multirow{2}{*}{x(\textbf{60})} & \multirow{2}{*}{x(\textbf{120})} \\
        $s_{2}^{y}=-0.01$, $s_{2}^{z}=-0.2$ &                                    &                                 &                                 &                                  \\
        \hline
      \end{tabular}
    \end{adjustbox}
    \caption{Benchmarks and speedups of v3\_Opt and v3\_Opt\_rk4 compared to
      v3.}\label{benchmarkresults}
  \end{table}

\begin{figure}
  \centering
  \begin{subfigure}[t]{0.5\textwidth}
    \centering
    \includegraphics[height=5cm]{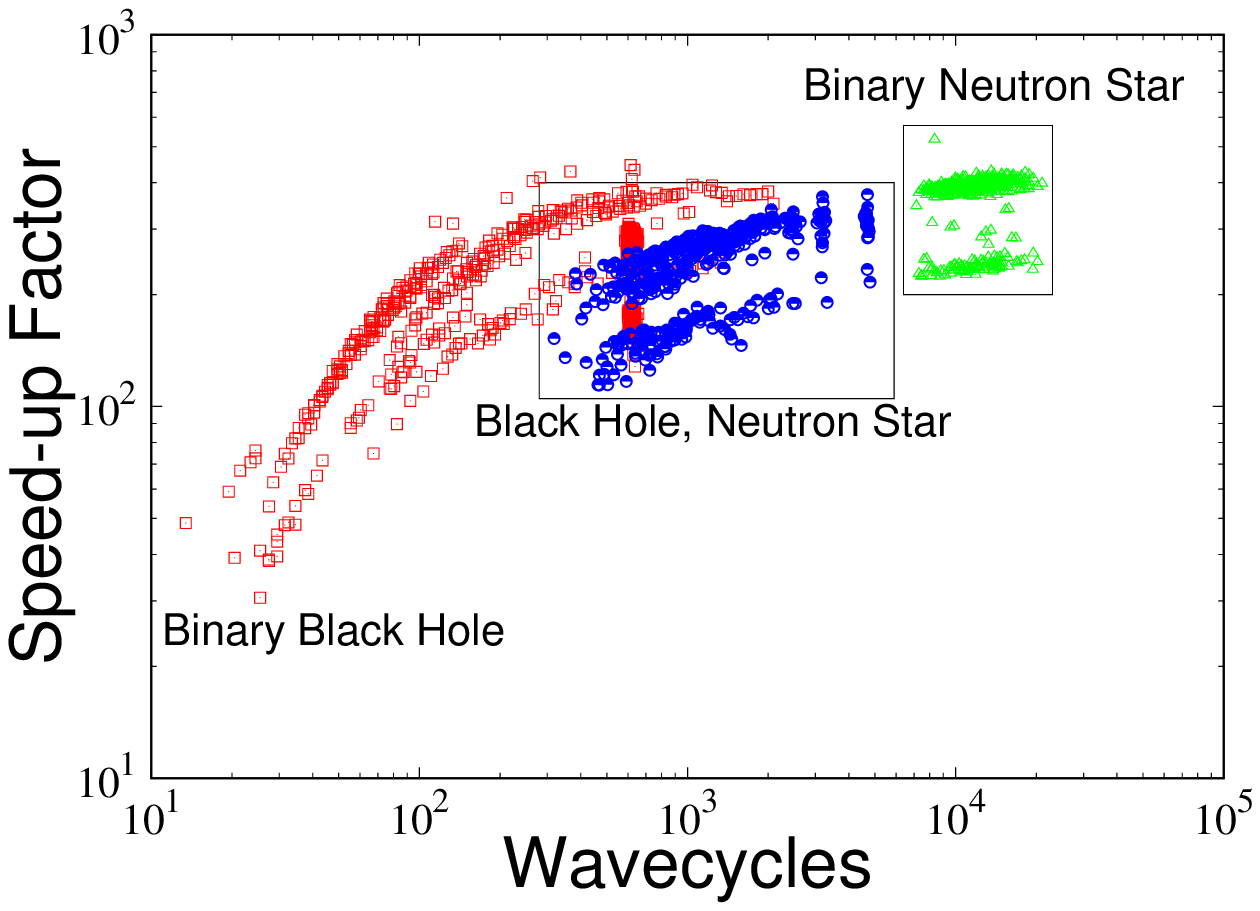}
  \end{subfigure}%
~
  \begin{subfigure}[t]{0.5\textwidth}
    \centering
    \includegraphics[height=5cm]{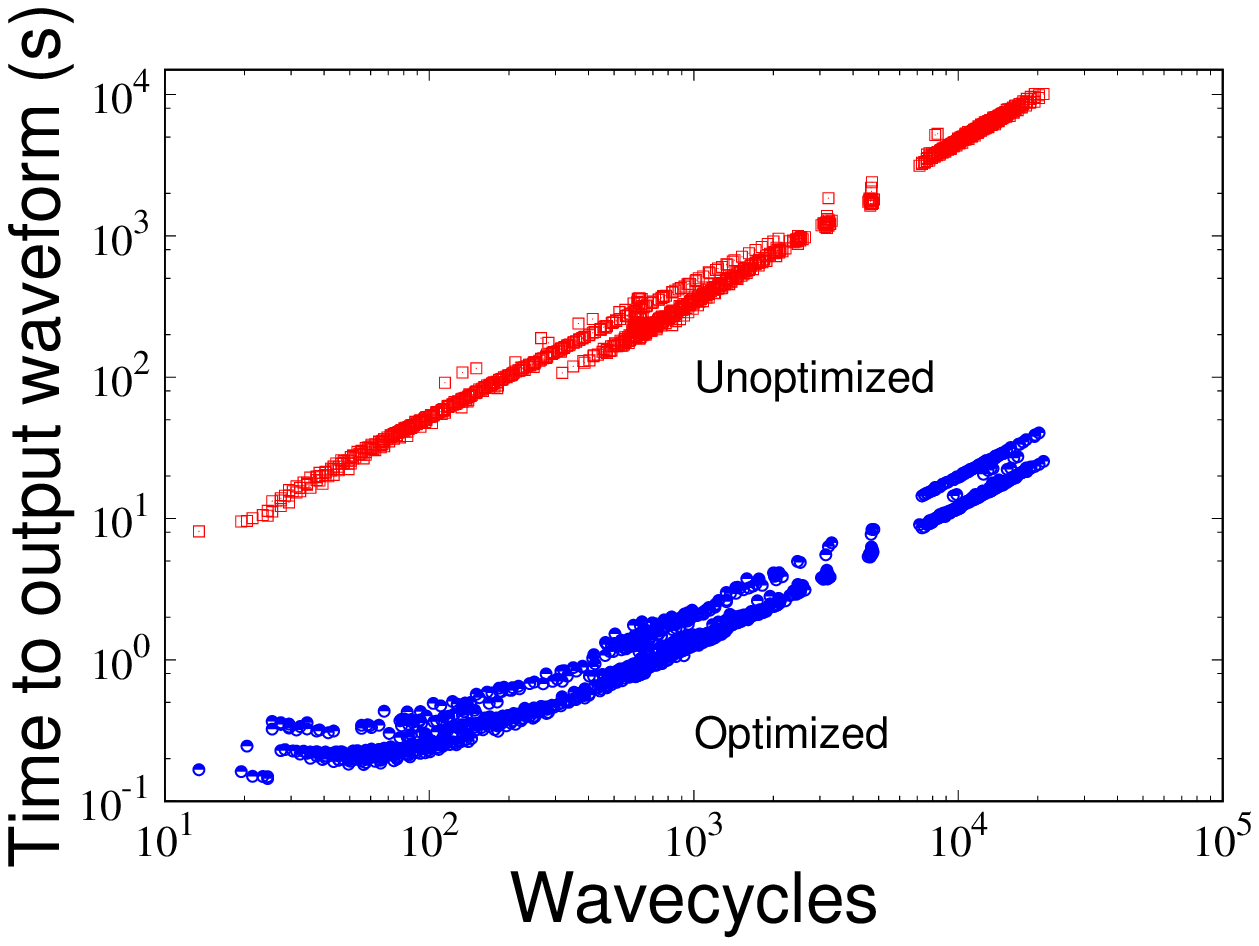}
  \end{subfigure}
  \caption{\textbf{Performance benchmarks:} \textbf{Left panel:} plots speedup
    factor versus number of wavecycles in the binary inspiral. Measuring the
    number of wavecycles allows us to compactly display the results of the
    benchmark tests without explicit reference to mass or spin. \textbf{Right
      panel:} plots the number of wavecycles versus the time taken to output
    the waveform. Note that the speedup factor in the left panel is simply the
    ratio of the curves in the right panel.}\label{speed}
\end{figure}

To demonstrate that the advertised speedup factors of Table
\ref{benchmarkresults} apply across the parameter space of binaries of
interest to the LVC, we completed four benchmark surveys. The first two
concern binary black hole systems, one with varying masses and the other with
varying spins. The third survey considers mixed binaries (one black hole and
one neutron star), and the fourth binary neutron stars. The parameters tested
in each run are included in Table \ref{parambench}. The results of these
surveys are plotted in Figure \ref{speed} and summarized in Table
\ref{benchmarkresults}.

  \begin{table}
    \begin{adjustbox}{max width=\textwidth}
    \begin{tabular}{lcccc}
      \hline
      Ranges        & $m_{1}$ ($\sunmass$) & $q$ (dimensionless)       & $a_{1}$ (dimensionless) & $a_{2}$ (dimensionless) \\
      \hline\hline
      BHB$_{\rm M}$ & $[16.7,100.3]$       & $[1,10]$                  & 0.0500001               & 0                       \\
      BHB$_{\rm S}$ & 10                   & 1                         & $[-0.95,0.95]$          & $[-0.95,0.95]$          \\
      BHNS          & $[7,100]$            & $\frac{M}{1.4}$           & $[-0.95,0.95]$          & 0                       \\
      DNS           & $[1.2,2.3]$          & $\frac{M}{m\in[1.2,2.3]}$ & 0.0500001               & 0                       \\
      \hline
    \end{tabular}
    \end{adjustbox}
    \caption{Surveyed parameters: each survey tested 400 parameter
      combinations, with 20 evenly-spaced values taken in each range
      indicated. Here BHB$_{\rm M}$ indicates the black hole binary mass
      survey, BHB$_{\rm S}$ the black hole binary spin survey, BHNS the black
      hole neutron star survey, and DNS the double neutron star survey. We
      define $q \equiv \frac{m_{1}}{m_{2}}$, the ratio of the mass of object 1
      to the mass of object 2. The dimensionless Kerr spins of each object are
      denoted $a_{1}$ and $a_{2}$, respectively. Each waveform generation
      started with a frequency of 10 Hz used a sample rate of 16,384
      Hz.}\label{parambench}
  \end{table}

We would like to measure an average speedup based on the four surveys. As
in~\cite{v2opt}, we define an overall speedup factor as a waveform
cycle-weighted average
  \[
    \mathcal{S} = \frac{\sum_{i}\mathcal{S}_{i}N_{i}}{\sum_{i} N_{i}}
  \]
where $\mathcal{S}_{i}$ is the speedup factor for generating the
$i^{\text{th}}$ waveform and $N_{i}$ is the number of
wavecycles in the $i^{\text{th}}$ waveform. We found $\mathcal{S}\sim340$.
This reduces the time necessary for a black hole binary PE run from
$\mysim$100 years (with v3\_preopt) to $\mysim$8 months (with v3\_Opt). We
expect lower mass PE runs will be possible on similar timescales with
additional optimizations.

\section{Conclusions and Future Work}
\label{Conclusions}
Anticipating the potential detection by Advanced LIGO of significantly
precessing compact binaries, we have optimized v3 to make costly
precessing-waveform-approximant-based data analysis applications like PE
possible in a reasonable amount of time. If an efficient 8D ROM is found, such
optimizations will make the construction of this ROM faster. After migrating
v2/v4 optimizations to v3, we further optimized partial derivatives of the
Hamiltonian using a GAD scheme. This resulted in waveforms  that are faithful
to v3, as evidenced by faithfulness increasing to 1 as ODE tolerance
decreases. We achieved an average overall speedup of $\mysim340$x, ranging
from $\mysim120$x for GW150914-like black hole binaries to $\mysim630$x for
black hole-neutron star binaries. We expect that further optimizations are
possible, achieving an additional speedup factor of at least $\mysim$3x.
Future work will focus on transforming Cartesian coordinates to spherical
coordinates to lower sampling rates even more during ODE solving and
integration.

\ack
We thank O.~Birnholtz, N.~Johnson-McDaniel, R.~Sturani, A.~Taracchini, and
C.~Haster for helpful comments and discussion during a review of the v3\_opt
code. A.~Taracchini is especially thanked for introducing us to faithfulness
testing via \texttt{PyCBC}, as is S.~Teukolsky for suggesting dense output
methods. We also thank R.~Haas for his work optimizing v3 during its
development, and I.~Ruchlin for numerous helpful discussions. This work was
supported primarily by NSF LIGO Research Support Grant PHY-1607405. Early work
on this project was supported by NSF EPSCoR Grant 1458952 and NASA Grant
13-ATP13-0077. We are grateful for the computational resources provided by the
Leonard E.~Parker Center for Gravitation, Cosmology and Astrophysics at
University of Wisconsin-Milwaukee (NSF Grant 0923409), LIGO Livingston
Observatory, LIGO Hanford Observatory, and the LIGO-Caltech Computing Cluster.

\section*{References}

\sloppy

\bibliographystyle{unsrt}
\bibliography{references}{}

\end{document}